\documentclass[aps,prl,twocolumn,superscriptaddress,superscriptaddress,10pt]{revtex4-1}

\usepackage{graphicx}
\usepackage{multirow}
\usepackage{rotating}
\usepackage{float}
\usepackage{epstopdf}
\usepackage{natbib}
\usepackage{amsmath}
\usepackage[utf8]{inputenc}
\usepackage{mathcomp}

\DeclareGraphicsExtensions{.pdf,.eps,.png,.jpg,.mps}

\begin{document}

\title{Investigation of the strongly correlated one-dimensional magnetic behavior of NiTa$_2$O$_6$}

\author{J. M. Law}
\email[]{j.law@hzdr.de}
\affiliation{Dresden High Magnetic Field Laboratory (HLD), Helmholtz-Zentrum Dresden-Rossendorf, D-01314 Dresden, Germany}

\author{H.-J. Koo}
\affiliation{Department of Chemistry and Research Institute of Basic Science, Kyung
 Hee University, Seoul 130-701, Korea}

\author{M.-H. Whangbo}
\affiliation{Department of Chemistry, North Carolina State
University, Raleigh, North Carolina 27695-8204, U.S.A.}

\author{E. Br\"ucher}
\affiliation{Max-Planck-Institut f\"ur Festk\"orperforschung,
Heisenbergstr. 1, D-70569 Stuttgart, Germany}

\author{V. Pomjakushin}
\affiliation{Laboratory for Neutron Scattering, Paul Scherrer Institut, CH-5232 Villigen, Switzerland}

\author{R. K. Kremer}
\affiliation{Max-Planck-Institut f\"ur Festk\"orperforschung,
Heisenbergstr. 1, D-70569 Stuttgart, Germany}

\date{\today}

\begin{abstract}
The magnetic properties of NiTa$_2$O$_6$ were investigated by magnetic susceptibility, specific heat, electron paramagnetic resonance, neutron powder diffraction and pulse field magnetization measurements. Accompanying \textit{ab~initio} DFT calculations of the spin-exchange constants complemented and supported our experimental findings that NiTa$_2$O$_6$ must be described as a quasi-1D Heisenberg $S$~=~1 spin chain system with a nearest-neighbor only anti-ferromagnetic spin-exchange interaction of 18.92(2)~K. Inter-chain coupling is by about two orders of magnitude smaller. Electron paramagnetic resonance measurements on Mg$_{1-x}$Ni$_x$Ta$_2$O$_6$ ($x \approx$ 1\%) polycrystalline samples enabled us to estimate the single-ion zero-field splitting of the  $S$~=~1 states which amounts to less than 4\%  of the nearest-neighbor spin-exchange interaction.
At 0~T NiTa$_2$O$_6$ undergoes long-range anti-ferromagnetic ordering at 10.3(1)~K evidenced by a $\lambda$-type anomaly in the specific heat capacity. On application of a magnetic field the specific heat anomaly is smeared out. We confirmed the magnetic structure by neutron powder diffraction measurements and at 2.00(1)~K refined a magnetic moment of 1.93(5)~$\mu_{\rm{B}}$ per Ni$^{2+}$ ion.  Additionally, we followed the magnetic order parameter as a function of temperature. Lastly we found saturation of the magnetic moment at 55.5(5)~T with a $g$-factor of 2.14(1), with an additional high field phase above 12.8(1)~T. The onset of the new high field phase is not greatly effected by temperature, but rather smears out as one approaches the long-range ordering temperature.
\end{abstract}


\maketitle

\section{Introduction}
The magnetic properties of one-dimensional (1D) spin chain systems have attracted special attention because they may realize exotic ground states due to the interplay of charge, spin and orbital degrees of freedom giving rise to partly complex excitations, which are far from being fully understood.\cite{Auciello1998,Cheong2007,Brink2008,Khomskii2009,Mourigal2011} In recent years we have identified and investigated the properties of new low-dimensional magnetic quantum antiferromagnets (AFM) which realize magnetic frustration along the chains due to a competition of nearest (nn) and next-nearest neighbor (nnn) spin-exchange interaction (SEI), the latter being mediated via two anions like e.g. O$^{2-}$ or Cl$^-$ or Br$^{-}$. \cite{Gibson2004,Enderle2005,Banks2009,Enderle2010,Law2010,Law2011,Law2011a,Lebernegg2013} 

Low-dimensional Ni$^{2+}$ compounds lately have attracted special attention because they constitute $S$~=~1 (3d$^8$ electronic configuration) systems. Ni$^{2+}$ linear chain compounds were found to be of particular interest because they can realize $S$~=~1 Haldane systems with a gap in the excitation spectra.\cite{Haldane1983,Haldane1983a,Affleck1989,Buyers1986,Steiner1987,Morra1988,Darriet1993}

\begin{figure}[H]
  \includegraphics[width=8cm]{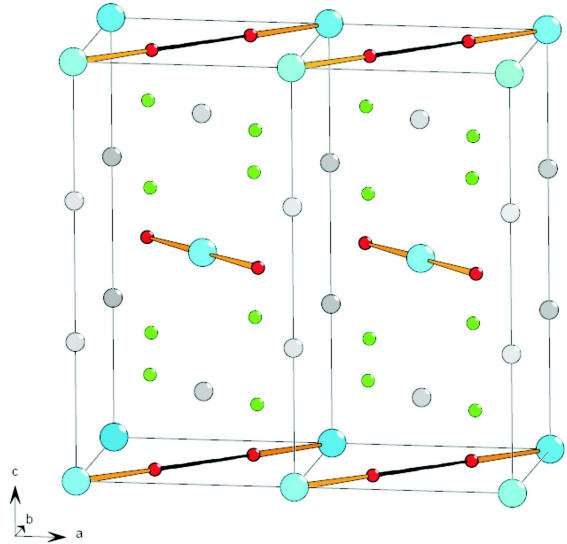}
  \caption{The crystal structure of NiTa$_2$O$_6$. Two unit cells are shown. The (cyan) large spheres represent Ni atoms, the (grey) medium spheres Ta atoms and the small (red, green) spheres the O atoms. The Ni $\cdots$ O $\cdots$  O $\cdots$  Ni bonds highlighted by (yellow, black) solid lines.}\label{chemstructure}
\end{figure}

NiTa$_2$O$_6$ is a chemically well characterized material that crystallizes in the tri-rutile structure type, which derives from the well-known rutile type as a consequence of the chemical ordering of the divalent and the pentavalent cations, Ni$^{2+}$ and Ta$^{5+}$, leading to a tetragonal structure with the $c$-axis being tripled as compared to $a$ and $b$ axes (see Figure \ref{chemstructure}). Apart from small orthorhombic distortions, the Ni and Ta atoms are octahedrally coordinated by oxygen atoms. The magnetic lattice consists of Ni$^{2+}$ ions occupying a body-centered tetragonal arrangement  resulting in square-planar Ni layers stacked along the $c$-axis.

The magnetic properties of NiTa$_2$O$_6$ have been the subject of a number of studies, but a unanimous consensus especially on the appropriate spin-exchange model has not been reached until now.\cite{Bernier1968,
Takano1970,KREMER1988,KREMER1988a,Ehrenberg1998,White1991,OliveiraNeto2007,Santos2010,OliveiraNeto2012,Santos2012}
Previous magnetization measurements have reported Curie-Weiss behavior with Curie-Weiss temperatures ranging from -19.9 K to -50~K indicating a predominant AFM SEI.\cite{OliveiraNeto2012,Takano1970,Bernier1968} Long-range AFM ordering was consistently reported to appear below 11~K.\cite{KREMER1988,KREMER1988a,Santos2010,Santos2012} The magnetic structure has been previously determined by powder neutron diffraction revealing a rather large magnetic unit cell  (propagation vector (1/4, -1/4, 1/2)) with the magnetic moments being collinearly aligned parallel to [1 1 0].\cite{Ehrenberg1998} The ordered moment was refined to 1.6 $\mu_{\rm B}$ and the difference to the expected moment of $g \times m_S \sim$ 2 $\mu_{\rm B}$ was attributed to incomplete ordering of the spins.

In a recent publication, Santos \textit{et al.} described NiTa$_2$O$_6$ as a two-dimensional (2D) AFM system. Their analysis was based upon a fit of the high-temperature magnetic susceptibility considering data above the AFM short-range ordering maximum. Their model included SEI along the edges of the squares and across the diagonal, given as $\cal{H}=\rm{-2}\textit{J}\sum{\textit{S}_i\textit{S}_j}$ and also included a single ion anisotropy term D, where D is defined as $\cal{H}=\rm -D\sum{\textit{S}_z^2}$ and an anisotropic $g$-factor. They found that both SEI were roughly equal and AFM with a value of ~3.4~K, D was found to be 56.7~K and an average $g$-factor of 3.08 (where $g_\parallel$~=~3.51 and $g_\perp$~=~2.03).\cite{Santos2012} Such a large single ion anisotropy implies that NiTa$_2$O$_6$ is close to an Ising-like system. However, the $g$-factors deviating so greatly from the free-electron $g$-factor, $g_{\rm{e}}$, are unprecedented for Ni$^{2+}$ in an octahedral oxygen environment.\cite{Bleaney1970} In addition, for an Ising-like system, $g_\perp$ is expected to be close to zero. 
In contrast, our recent re-analysis of the magnetic susceptibility of NiTa$_2$O$_6$ using Pad\'{e} approximation of Quantum Monte Carlo calculations showed that a one-dimensional $S$ = 1 Heisenberg spin chain scenario is appropriate to describe the magnetism of NiTa$_2$O$_6$.\cite{Law2013} 
In order to resolve this discrepancy we have carried out a complete re-analysis of the magnetic properties of NiTa$_2$O$_2$.

Our work is organized as follows: Firstly we describe the results of density functional (DFT) calculations performed in order to evaluate the spin-exchange constants appropriate for NiTa$_2$O$_6$. In a second part we report electron paramagnetic resonance (EPR) measurements on Ni$^{2+}$ ions doped into the isostructural diamagnetic matrix in MgTa$_2$O$_6$ carried out so as to evaluate the single-ion properties especially the zero-field splitting of the $S$ = 1 manifold. We re-analyze the magnetic susceptibility and the temperature and magnetic field dependence of the heat capacity and propose a magnetic phase diagram. Finally we re-determine the magnetic structure, which enables us to drive the ordered magnetic moment. Our analysis unequivocally proves that NiTa$_2$O$_6$ represents a $S$ =1 Heisenberg chain with AFM nn SEI along the [110] direction.

\section{Density Functional Calculations of the Spin-Exchange}

\begin{figure}[h]
 \centering
  \includegraphics[width=4cm]{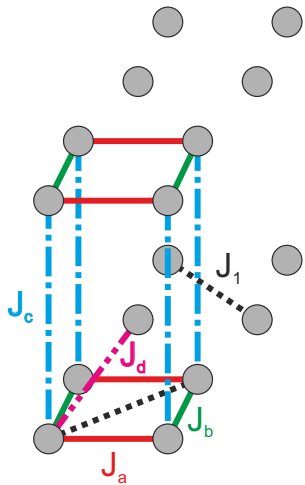}
  \caption{(Color Online) The spin-exchange pathways used for the DFT calculations.}\label{DFTbonds}
\end{figure}

In order to investigate the spin-exchange of NiTa$_2$O$_6$, we consider the five spin-exchange paths defined in Figure \ref{DFTbonds}. J$_{a,b,c}$ are the SEI along the respective axes from one corner to another, J$_d$ is the predominant inter-plane coupling while J$_1$ is the predominant intra-plane coupling. An intra-plane SEI perpendicular to J$_1$ was neglected since previous H\"{u}ckel-extended tight binding calculations have indicated it to be small.\cite{Koo}

\begin{figure}[H]
  \includegraphics[width=8cm]{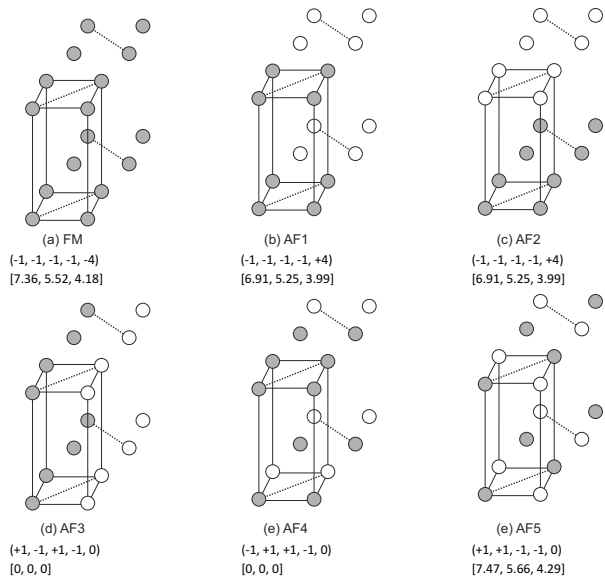}
  \caption{ Six ordered spin states constructed by using a (2a, 2b, 2c) supercell containing 16 FUs, where the filled and unfilled circles represent up-spin and down-spin Ni$^{2+}$ sites, respectively. The five numbers in each parenthesis, from left to right, are the coefficients n$_{\textit{1}}$, n$_{\textit{a}}$, n$_{\textit{b}}$, n$_{\textit{c}}$, and n$_{\textit{d}}$, of Eq. (1), which determine the total SEI energy per FU, and the three numbers in each square bracket, from left to right, represent the relative energies (meV/FU) determined from the GGA+U calculations with U$_{\rm{eff}}$ = 3, 4 and 5 eV, respectively. }\label{spinarrangements}
\end{figure}

To determine the energies of the five SEI, we examined the relative energies of the six ordered spin states depicted in Figure \ref{spinarrangements} in terms of the Heisenberg spin Hamiltonian;
\begin{equation}
\cal{H}=-\rm \sum{\textit{J}_{ij}\vec{\textit{S}_i}\vec{\textit{S}_j}},
\end{equation}
where \textit{J}$_{ij}$ is the exchange parameter for the coupling between spin sites \textit{i} and \textit{j}. Then, by applying the energy expressions obtained for spin dimers with N unpaired spins per spin site, see Figure \ref{spinarrangements}, \cite{Dai2001,Dai2003} the total spin-exchange energies of the six ordered spin states, per formula unit (FU), can be expressed in the following form 
		\begin{equation}
		E_{\rm{FU}}=(n_{\textit{1}} J_{\textit{1}} + n_a J_a + n_b J_b + n_c J_c + n_d J_d) (N^2/4),
		\label{eq:SEI}
		\end{equation}
		where n$_i$ ( \textit{i} = \textit{1}, \textit{a}, \textit{b}, \textit{c}, \textit{d}) refers to the coefficient of the spin exchange J$_i$. These coefficients for the six ordered states are summarized in Figure \ref{spinarrangements}. 
We determine the relative energies of the six ordered spin states of NiTa$_2$O$_6$ on the basis of DFT calculations using the Vienna \textit{ab~initio} simulation package, employing the projected augmented-wave method, the generalized
gradient approximation (GGA) for the exchange and correlation functional, with the plane-wave cut-off energy set to 400 eV, and a set of 30 k-points for the irreducible Brillouin zone.\cite{Kresse1993,Kresse1996,Kresse1996a,Perdew1996} To account for the strong electron correlation associated with the Ni 3$d$ state, we performed GGA plus on-site repulsion (GGA+$U$) calculations with $U_{\rm eff}$ = 3, 4 and 5 eV for Ni.\cite{Dudarev1998} The relative energies of the six ordered spin states obtained from our GGA+$U$ calculations are summarized in Figure \ref{spinarrangements}. Then, by mapping these relative energies onto the corresponding relative energies from the total spin-exchange energies, equation \ref{eq:SEI}, \cite{Whangbo2003,Koo2008,Koo2008a,Kang2009,Koo2010} we obtain the values of the spin-exchange parameters as summarized in Table \ref{table:SEI}.

\begin{table}[h]
	\centering
		\begin{tabular}{c|c|c|c}
		SEI	& $U_{\rm{eff}}$ = 3 eV	&	$U_{\rm{eff}}$ = 4 eV	&	 $U_{\rm{eff}}$ = 5 eV\\
		\hline
		\hline
J$_{\textit{1}}$&-42&-32&-24\\
J$_a$&1.0&0.76&0.59\\
J$_b$&1.0&0.76&0.59\\
J$_c$&-0.25&-0.18&-0.16\\
J$_d$&-0.65&-0.42&-0.26\\
		\end{tabular}
		\caption{The spin-exchange parameters (in K) obtained from GGA+$U$ calculations.}\label{table:SEI}
\end{table}

\textit{J}$_1$ is the dominant SEI (Table \ref{table:SEI}) exceeding the other SEI by two orders of magnitudet, which supports our experimental findings (see below) that NiTa$_2$O$_6$ constitutes a nn $S$~=~1 spin chain with AFM nn SEI.

\section{Experimental}
\subsection{Sample preparation}

Powder samples of NiTa$_2$O$_6$ and Mg$_{1-x}$Ni$_x$Ta$_2$O$_6$ ($x$ = 0.001 and 0.01) were prepared, in accordance with Takano \textit{et al.}, by mixing NiO or MgO and Ta$_2$O$_5$ (Alfa Aesar, all materials Puratronic) in stoichiometric quantities and heating to 1300 $^{\rm o}$C for 48 hours.\cite{Takano1970} Multiple  re-grindings and repetition of the annealing process and  X-ray powder diffraction was performed and additional starting materials were added, if needed, until phase purity was reached. 

\subsection{Magnetic Susceptibility, Magnetization and Specific Heat}

Magnetic susceptibility of a 145~mg powder sample of NiTa$_2$O$_6$ were measured with a SQUID magnetometer (MPMS XL, Quantum Design). Pulse field isothermal magnetization up to $\sim$~60~T were measured at the Hochfeld-Magnetlabor Dresden, Helmholtz-Zentrum Dresden-Rossendorf, Germany on a 35.4~mg sample, using compensated coils.\cite{Skourski2011} In order to determine the scale factor for the pulse field results magnetization up to 14~T of the identical sample were also determined using the VSM option of a physical property measurement system (PPMS) (Quantum Design).

The specific heat was measured on a 76~mg pelletized sample using a PPMS at various fields between 0 and 14~T.

\subsection{Electron Paramagnetic Resonance}
EPR spectra were collected at $\sim$ 9.4 GHz with a Bruker ER 040XK X-band spectrometer in an ER73 electromagnet controlled by a B–H-015 field controller which was calibrated against the resonance of 2,2-diphenyl-1-picrylhydrazyl (DPPH).

\section{Results and Discussion}

\subsection{Zero-field splitting}

A dominant feature of the magnetism of Ni$^{2+}$ in a nearly octahedral environment is the zero-field splitting of the ground spin triplet state ($\Gamma^2 $, $S$ = 1). In order to determine the magnitude of the single-ion zero-field splitting of the Ni$^{2+}$ ions in NiTa$_2$O6 we measured the electron paramagnetic resonance of highly diluted Ni$^{2+}$ entities in  the diamagnetic matrix MgTa$_2$O$_6$.
MgTa$_2$O$_6$ is isostructural with NiTa$_2$O$_6$ with lattice parameters $a$ = 4.7189(7) \AA~and $c$ = 9.2003(22) \AA, only slightly different from those of NiTa$_2$O$_6$ ($a$ = 4.7219(11) \AA~and $c$ = 9.150(5) \AA), and with nearly identical position parameters of the oxygen atoms within error bars.\cite{Mullerbuschbaum1986,Halle1988} 
\begin{figure}[H]
  \includegraphics[width=8cm]{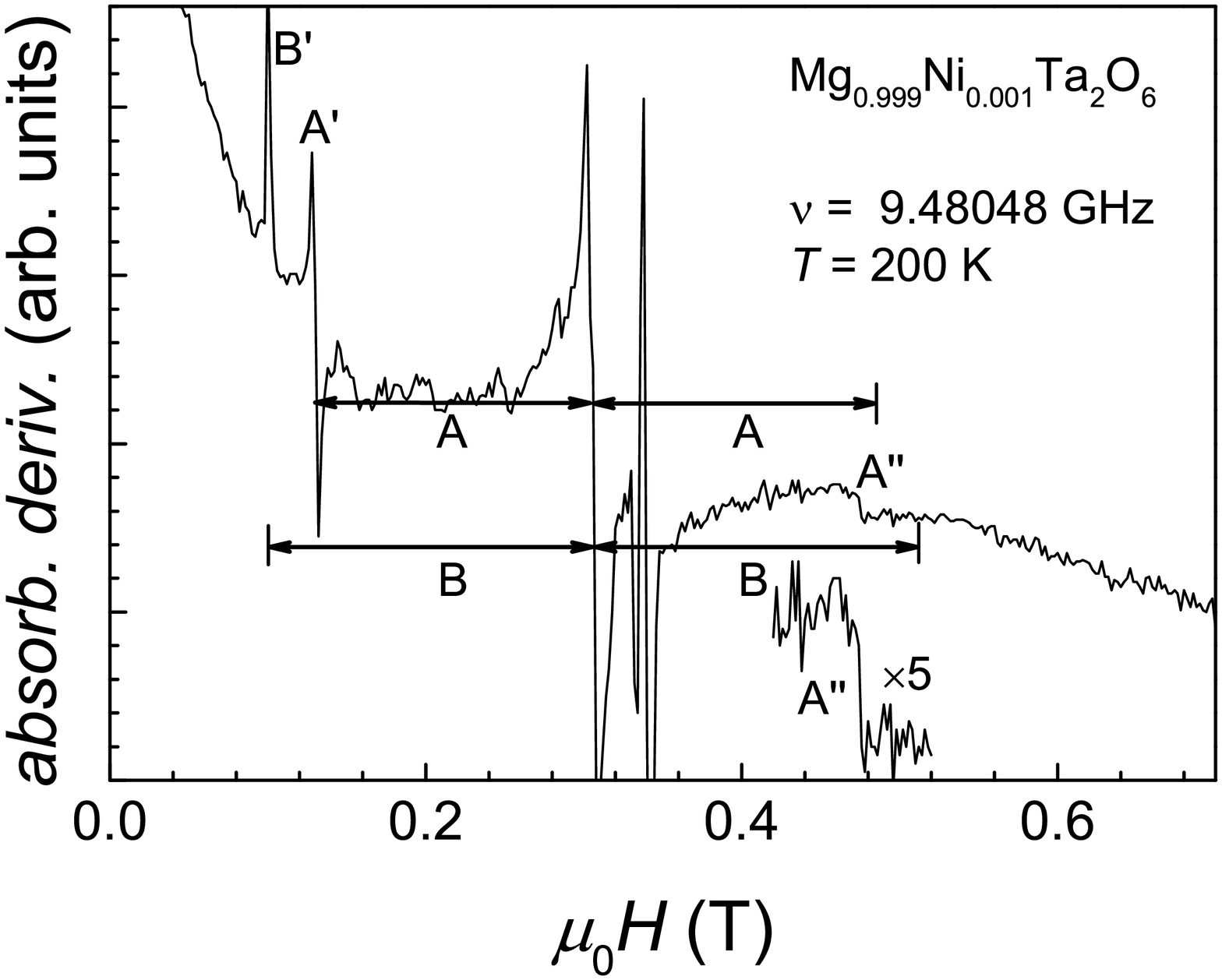}
  \caption{EPR spectrum of polycrystalline Mg$_{0.999}$Ni$_{0.001}$Ta$_2$O$_6$ measured at 200 K with microwave frequency of 9.4805 GHz. The vertical bars A and B mark the distances of the satellites from a center field of 0.306 T corresponding to a $g$-factor of 2.21. A section of the spectrum near 0.47 T has been magnified by a factor of 5 as indicated.}\label{eprspectrum}
\end{figure}

Figure \ref{eprspectrum} displays an EPR spectrum of Mg$_{0.999}$Ni$_{0.001}$Ta$_2$O$_6$ collected at 200~K. It exhibits two intensive resonance lines at $g$-factors of 2.21 (resonance field 0.3068~T) and 2.00 (resonance field 0.3390~T) and two less intensive satellites at 0.1002~T (B\textquotesingle) and 0.1303~T (A\textquotesingle). On the high field side a very weak resonance at 0.4752~T (A\textquotesingle\textquotesingle) becomes visible on magnification. The latter two satellites have the same distance, A = 0.173~T, from the resonance at 2.21, while a satellite, B\textquotesingle\textquotesingle~at high fields, symmetric to B\textquotesingle ~at the low-field side could not be detected. The satellites A\textquotesingle~, A\textquotesingle\textquotesingle ~and B\textquotesingle ~have similarly been seen in a sample of Mg$_{0.99}$Ni$_{0.01}$Ta$_2$O$_6$. The linewidth of the $g$ = 2.21 line is considerably larger and the narrow $g$~=~2.0 resonance line is hidden by the broad $g$~=~2.21 line. Q-band measurements taken at $\sim$~34~GHz show similar spectra, however shifted to higher fields corresponding to the larger microwave frequency.\cite{Bruker}

The EPR spectra of  Mg$_{1-x}$Ni$_{x}$Ta$_2$O$_6$, particularly shape and position of the satellites with respect to a central resonance line at $g$ = 2.21,  are reminiscent of the EPR spectra of randomly oriented triplet systems with zero-field splitting of the threefold degenerate state due to a crystal field of symmetry lower than axial.

The spin Hamiltonian, $H$ with the $z$ axis chosen as the unique axis of the crystal field invoking an axial and a rhombic crystal field is given by

\begin{equation}
\cal{H} = \rm H_{\rm Zee} + D [S_z^2 - \frac{1}{3} S(S + 1)] + E (S_+^2 + S_-^2),
\label{Eq:ham}
\end{equation}

where \textit{D} and \textit{E} are the axial and rhombic crystal field parameters, respectively. They typically amount to a few Kelvin, with \textit{E} smaller than \textit{D}.\cite{Bleaney1970}

Wasserman \textit{et al.} have  calculated the EPR spectra of the triplet states of randomly oriented molecules and found three satellites symmetrically placed on the high- and low-field side with respect to the center resonance field. In the microwave absorption derivative the four outmost satellites have a positive amplitude while the two inner satellites have the shape of standard derivatives of EPR resonance lines with a positive and negative amplitude, indicating a peak in the direct EPR powder spectrum while the former outer satellites result from sharp edges in the EPR powder spectrum.\cite{Wasserman1964}
The four outmost satellites have a distance of $\lvert D\rvert$ and $\lvert D + 3 E\rvert $/2 with respect to the center, while the two inner satellites are by $\lvert D~–~E\rvert$/3 displaced from the center resonance field.\cite{Weil1994}

Applying this scenario to the EPR spectra of Ni$^{2+}$ in MgTa$_2$O$_6$ we can identify the two satellites (A\textquotesingle~and B\textquotesingle) at the low field side of the $g$ = 2.21 resonance line with the two innermost satellites of a random $S$ = 1 triplet. With this assignment and the relations given above we obtain for $D$ and $E$ at 200 K,

\begin{align*}
    D =& \pm 0.5025(50)~\rm{T},\\
    E =& \mp 0.031(1)~\rm{T}.
\end{align*}

D and E have opposite signs and differ in magnitude by a factor of $\sim$ 16. With a $g$-factor of $g$ = 2.2 these values correspond to, $D$ = $\pm$ 0.539 cm$^{-1}$ and $E$ = $\mp$ 0.014 cm$^{-1}$. The large $D$ value shifts the outmost satellite at the low field side out of our accessible field range thus allowing to understand why only two satellites can be detected.
A clear assignment of the signs require measurements at low temperature ($h \nu \approx k_{\rm B} T$).

The measured values of $D$, $E$ and the $g$-factor are in a range typically found for Ni$^{2+}$ in a slightly distorted octahedral environment and a lot more reasonable than what was concluded by Santos \textit{et al.}.\cite{Bleaney1970,Santos2012}

\begin{figure}[H]
  \includegraphics[width=8cm]{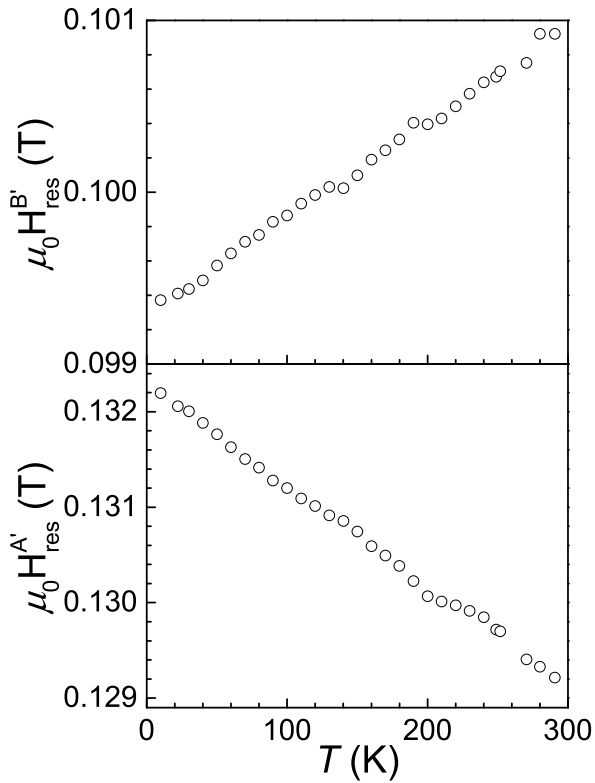}
  \caption{(o) The resonance fields, $\mu_0 H$ of the low-field resonance lines A\textquotesingle~and B\textquotesingle~versus temperature (see Figure \ref{eprspectrum}) measured with a microwave frequency of  9.480~GHz.}\label{eprtemp}
\end{figure}

Highly resolved temperature dependent measurements of the low-field part of the spectrum reveal small linear changes of the resonance positions of the resonance lines A\textquotesingle~and B\textquotesingle~(see Figure \ref{eprtemp}), which indicates a marginal temperature dependence of the crystal field parameters $D$ ($\approx$ 3 \%) and $E$ ($\approx$1.5 \%). This can be understood as due to thermal contraction of the lattice.

It is unexpected that only the low-field part of the spectrum can be detected, the high-field part being absent likewise in the X-band and Q-band spectra. The reason for this absence is not fully understood and cannot be attributed to straightforward transition-probability considerations.\cite{Orton1968}

The two rather strong resonance lines at $g$ = 2.21 and $g$ = 2 are not expected within the scope of the EPR of a randomly oriented triplet. At present we therefore tentatively assign them to magnetic defects ($g$ = 2) and/or small traces of magnetic impurities, e.g. other transition metals.

\section{Short-range correlation and long-range ordering}

Within the temperature and field range that was measured, no field dependence of the magnetic susceptibility was observed. This indicates that the as-prepared sample is rather clean with few ferromagnetic impurities. The high temperature ($\geq$ 150~K) magnetic susceptibility of NiTa$_2$O$_6$ can be explained by a Curie-Weiss behavior with a $g$-factor~=~2.16(1), a $\theta_{\rm{CW}}$~=~-32(1)~K and a temperature independent background $\chi_{\rm{0}}$~=~+83(3)~$\times$~10$^{-6}$ cm$^3$/mol (taken from the low-temperature fits, see below), see Figure \ref{sus}. These values are in agreement with previous findings (see above) and within the expected range for Ni$^{2+}$.

\begin{figure}[H]
  \includegraphics[width=9cm]{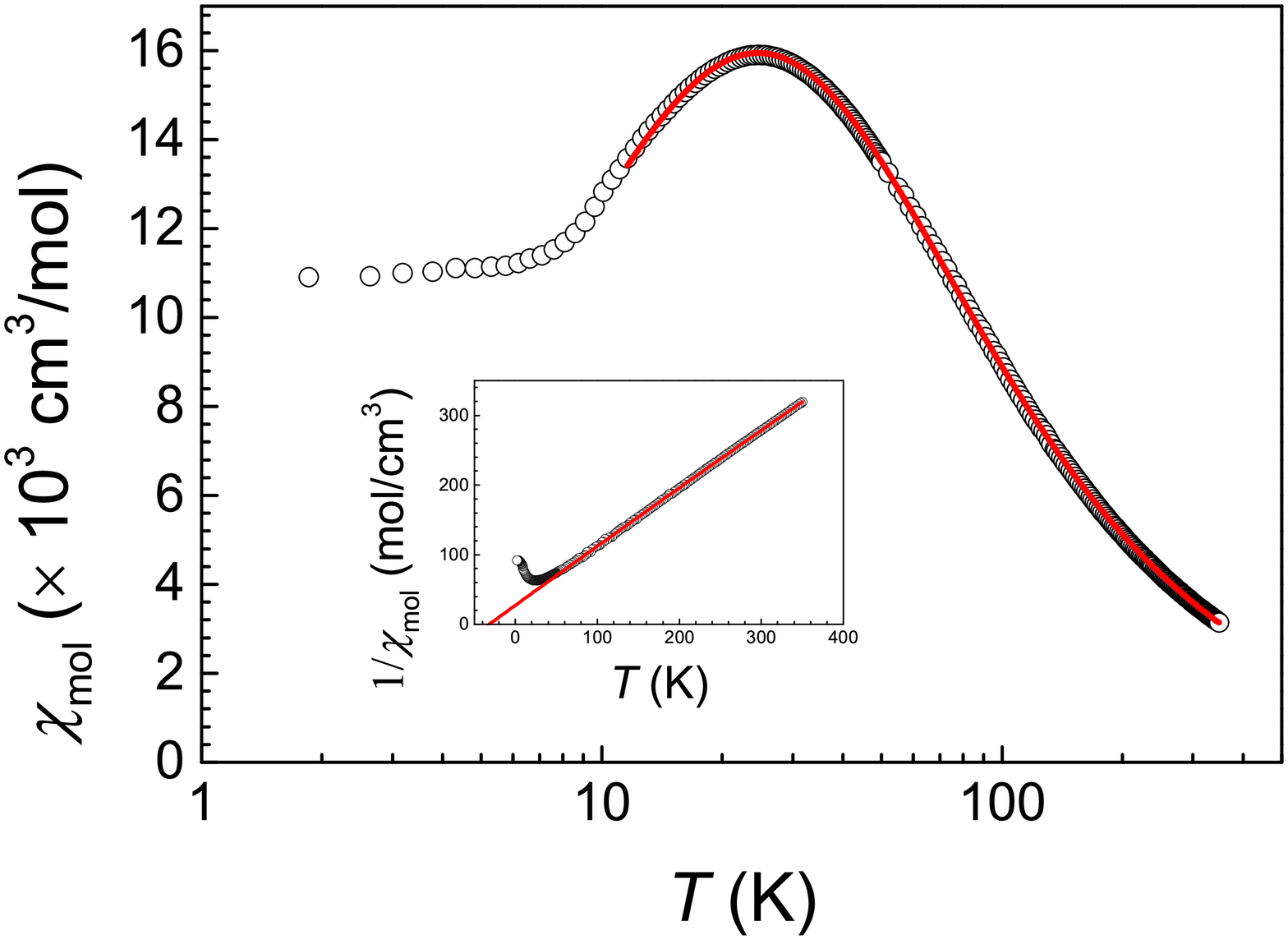}
  \caption{(Color online) Main: (o) The magnetic susceptibility of NiTa$_2$O$_6$, measured in a field of 7~T. Solid (red) line represents a fit to the data using the our Pad\'{e} approximation to the susceptibility of a $S$ =1 Heisenberg  AFM with nn SEI as described in the text. Inset: (o) The reciprocal magnetic susceptibility versus temperature. Solid (red) line is a Curie-Weiss fit to the data, see text for details.}\label{sus}
\end{figure}

The low-temperature magnetic susceptibility is dominated by a broad maximum, centered at $\approx$~24.5~K, which is indicative of low-dimensional AFM short-range ordering preceding the onset of long-range AFM ordering observed below $\sim$~11~K. Using the Pad\'{e} approximation for a $S$ = 1 Heisenberg chain with nn SEI, put forth by us previously, we can fit the broad maximum and subsequent high temperature ($\geq$ 11.5~K) magnetic susceptibility.\cite{Law2013} We added an additional term, $\chi_{\rm{0}}$ (a constant background term), to account for both the diamagnetic (negative) contribution of the closed shells of all atoms and a van Vleck (positive) contribution of the Ni$^{2+}$ atoms. Upon fitting the temperature dependent magnetic susceptibility, see Figure \ref{sus}, a value of 2.140(2) was found for the $g$-factor, which is reasonable for Ni$^{2+}$ in an octahedral crystal field coordination.\cite{Bleaney1970} The constant background term was fitted to $\chi_{\rm{0}}$~=~+83(3)~$\times$~10$^{-6}$ cm$^3$/mol, which is acceptable since the diamagnetic term using Selwood’s increments \cite{Selwood1956}  should be equal to \mbox{-112~$\times$~10$^{-6}$ cm$^3$/mol}, implying a van Vleck contribution of $\approx$~200~$\times$~10$^{-6}$ cm$^3$/mol which is reasonable for such an ion.\cite{Carlin1986} The nn SEI, $J_{\rm{nn}}$, converged to -18.92(2)~K, which is in agreement with the Curie-Weiss temperature found from the high temperature fit, since;

\begin{equation*}
\theta_{\rm{CW}}=\frac{S\times (S+1)}{3}\sum\limits_{i=1}^\infty\rm{z}_i~\rm{J}_i,
\end{equation*}

where $\theta_{\rm{CW}}$ is the Curie-Weiss temperature, z is given by the sum over the number of neighbors each ion has with the spin-exchange $J_i$. If one considers only the nn only SEI a Curie-Weiss temperature of $\approx$~-25.3~K is expected, the deviation from this value and the measured values can be attributed to the additional inter-chain SEI one finds in a quasi 1D system. Our fit is demonstrated in a semi-log plot in order to highlight any deviations between the data and the model, when compared to the fit of Santos \textit{et al.}, it is clear that we fully capture, very well, the maximum whereas their model was only applied at temperatures above. 
At lower temperature ($\approx$~11~K) there is a change of behavior in the temperature dependence of the magnetic susceptibility, indicative of the onset of long-range magnetic ordering. This is in agreement with other results presented herein and other results already published (see above).

\begin{figure}[H]
  \includegraphics[width=8cm]{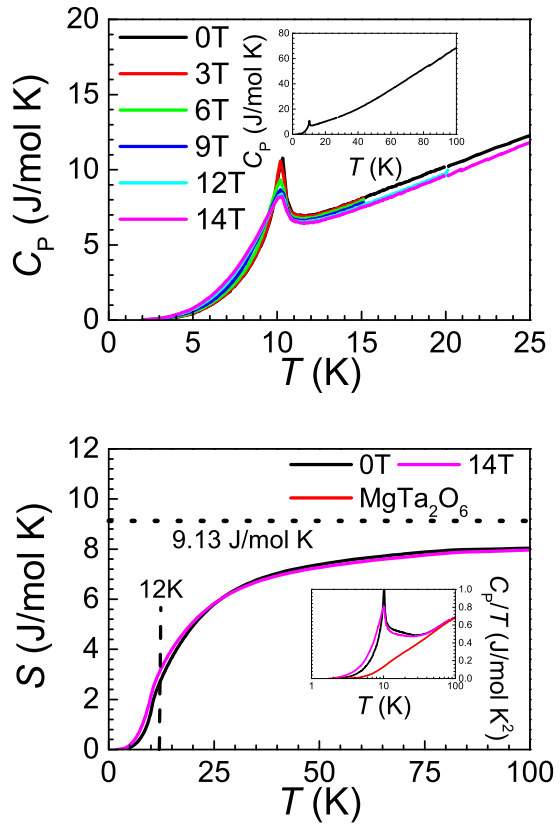}
  \caption{(Color online) Upper: The specific heat of NiTa$_2$O$_6$ versus temperature, measured in various magnetic fields. Upper inset: The specific heat versus temperature, measured at zero field over a larger temperature range. Lower: The entropy versus temperature for 0 and 14~T. Lower inset: $C_{\rm{P}}$/$T$ versus temperature for NiTa$_2$O$_6$ measured in both 0 and 14~T and MgTa$_2$O$_6$ measured in 0~T, see text for details.}\label{Cp}
\end{figure}

The specific heat shows clear evidence of long-range magnetic ordering at 10.3(1)~K as evidenced by a $\lambda$-type anomaly. With the application of a magnetic field the long-range ordering anomaly starts to smear out, but no clear evidence is seen for a shift in the onset, see Figure \ref{Cp}. We additionally measured the heat capacities of MgTa$_2$O$_6$ and used it as a reference for the lattice contributions to the heat capacity. In order to adjust for differences in the phonon spectrum of MgTa$_2$O$_6$ and NiTa$_2$O$_6$ the temperature axis of the specific heat of MgTa$_2$O$_6$ was uniformly compressed (0.92) such that the specific heats of both compounds matched at sufficiently high temperatures ($\geq$ 75~K). Then by integrating the difference i.e. the magnetic contribution \textit{C}$_{\rm{P}}$/\textit{T}, versus \textit{T} we can follow the magnetic entropy versus temperature. As can be seen in Figure \ref{Cp} approximately 73~\% of the total entropy of NiTa$_2$O$_6$ is contained in the specific heat in short-range correlation above the long-range ordering temperature. This is in agreement with results already published.\cite{KREMER1988a} A slight redistribution of the entropy is visible for the 14~T data.

We expect that a $S$~=~1 system should contain a total magnetic entropy of $R$ ln (2$S$+1) = 9.13~J/molK, but we only found 8.05~J/molK. The missing entropy can be attributed to the lattice contribution mismatch making it difficult to trace small magnetic contributions to the heat capacity especially at higher temperatures.

\section{Neutron diffraction}

\begin{figure}[H]
  \includegraphics[width=8cm]{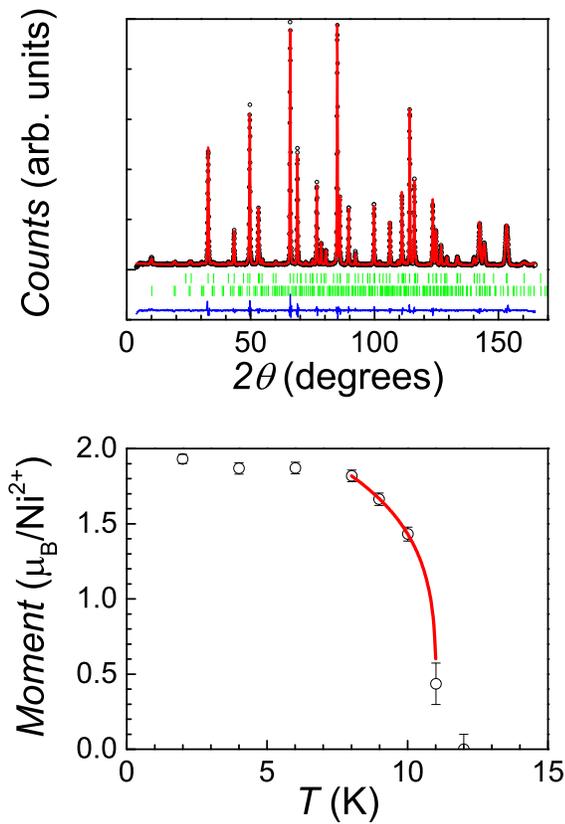}
  \caption{(Color online) Upper: The measured neutron diffraction pattern of NiTa$_2$O$_6$ at 1.99(1)~K (wavelength 1.8857 \AA) collected on HRPD, PSI Switzerland. Solid (red) line: Calculated pattern using the magnetic structure, see text for details. Solid (blue) line: Difference between measured and calculated patterns (offset). The positions of the magnetic Bragg reflections used to calculate the pattern are marked by the (green) vertical bars in the lower part of the Figure, the upper row contains the structural Bragg peaks and the lower row the magnetic Bragg peaks. Lower: (o) The refined magnetic moment versus temperature, the solid (red) line is a fit to a critical exponent, see text for details.}\label{neutrons}
\end{figure}

\begin{figure}[H]
  \includegraphics[width=8cm]{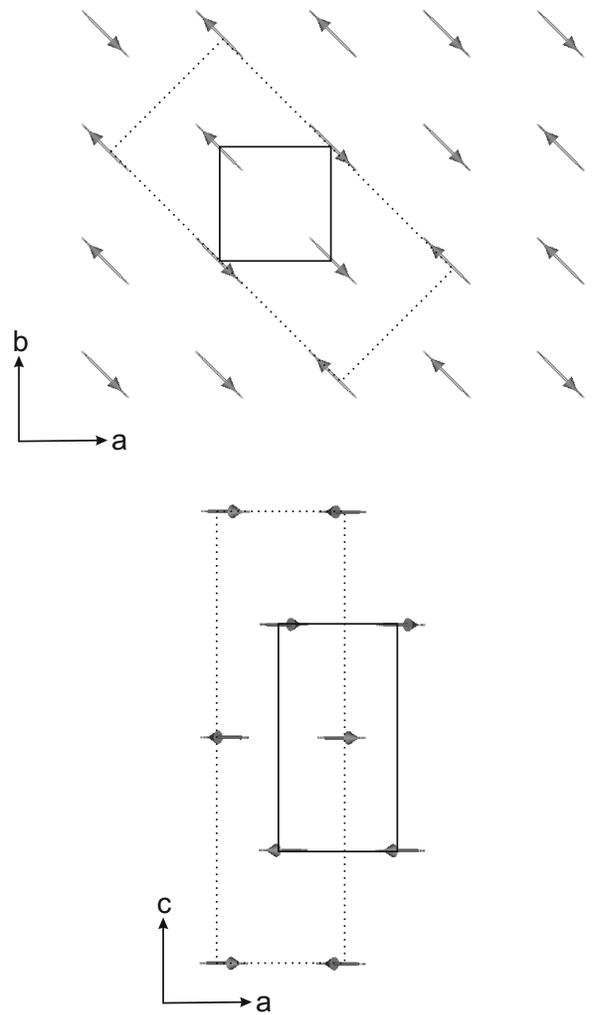}
  \caption{The magnetic structure of NiTa$_2$O$_6$ refined from the neutron powder diffraction pattern collected at 1.99~K. The solid (black) box represents the chemical unit cell and the dashed (black) box the magnetic unit cell. Upper: The magnetic moment arrangement of one layer look along the $c$-axis. Lower: Projection along the [010].}\label{structure}
\end{figure}

Powder neutron diffraction patterns were collected on the high resolution, medium intensity neutron diffractometer HRPT (PSI Switzerland) using neutrons with a wavelength of $\lambda$~=~1.8857~$\rm{\AA}$, at various temperatures between $\approx$ 2 and 20~K.\cite{HRPT} The powder diffraction pattern collected at 2 K is displayed in Figure \ref{neutrons}, the additional magnetic Bragg peaks were indexed on the basis of the magnetic propagation vector $\tau$~=~[$\frac{1}{4}$,$\frac{-1}{4}$,$\frac{1}{2}$], as found in the previous work.\cite{Ehrenberg1998} The  magnetic structure is shown in Figure \ref{structure}. At 2.00(1)~K we refined a magnetic moment of 1.93(5)~$\mu_{\rm{B}}$ per Ni$^{2+}$ ion  in good agreement with the expected value of 2~$\mu_{\rm B}$ for a $S$ = 1 system with a $g$-factor of 2.2. Upon increasing temperature the intensity of the magnetic Bragg peaks decreased until by 12.00(1)~K they vanished (see Figure \ref{neutrons}). Close to the long-range ordering temperature the magnetic moment can be fitted to a power law with a critical exponent, $\beta$ according to 

\begin{equation}
M(T) = M_0  \left( 1- T/T_{\rm{C}}\right) ^{\beta}, 
\end{equation}

We refined values of $\beta$~=~0.22(1) and $T_{\rm{C}}$~=~11.02(1). $T_{\rm{C}}$ is consistent with our other results. $\beta$ is smaller than expected for a Heisenberg system possibly due to the limited range of the reduced temperature.

The neutron diffraction supports the specific heat and magnetic susceptibility results that long-range ordering happens at 10.3(1)~K, the magnetic structure also confirms the DFT results and implies that the chain propagates along the [-1,-1,0] or [1,1,0] directions alternating along the $z$-axis.

\section{High magnetic field magnetization}

Pulse field isothermal magnetization up to $\sim$~60~T was measured at the Hochfeld-Magnetlabor Dresden, Helmholtz-Zentrum Dresden-Rossendorf, Germany on a 35.4~mg sample, using a compensated coil setup.\cite{Skourski2011} The identical sample was measured up to 14~T using the VSM option of the PPMS, in order to extract absolute values from the pulse field results.

\begin{figure}[H]
  \includegraphics[width=8cm]{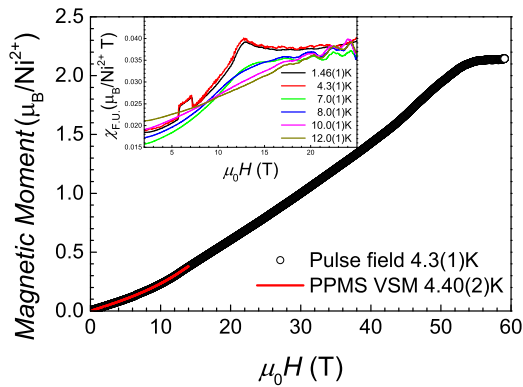}
  \caption{(Color online) Main: (o) Magnetization in $\mu_{\rm{B}}$ per Ni$^{\rm{2+}}$ atom measured in pulse field at 4.3(1)~K scaled to agree with the magnetization measured using a VSM, solid (red) line. Inset: The susceptibility measured in $\mu_{\rm{B}}$ per Ni$^{\rm{2+}}$ per T (taken from scaled pulse field measurements) versus magnetic field for various temperatures.}\label{pulsefield}
\end{figure}

The results of the magnetization measurements on a polycrystalline sample, determined in the pulse field magnetometer at 4.3(1)~K, can be seen in Figure \ref{pulsefield}. Saturation of the magnetic moment is observed at field above 55.5~T with a saturation moment of 2.14(1) $\mu_{\rm{B}}$ per Ni$^{\rm{2+}}$ ion. The saturation moment, in $\mu_{\rm{B}}$, is given by $gS$, where $g$ is the $g$-factor and $S$ is the spin of the ion. As such we find, from pulse field magnetization measurements, a $g$-factor of 2.14(1) which is in perfect agreement with that obtained from the analysis of the temperature-dependent magnetic susceptibility, see Figure \ref{sus}. This does not support the findings of Santos \textit{et al.}. \cite{Santos2012}

At lower fields we see clear evidence for a phase transition at 12.80(5)~T. With increasing temperature the transitions stay at the same field but start to become broader. At temperatures above the long-range ordering temperature the transition is no longer visible.

\subsection{Phase diagram}

\begin{figure}[H]
  \includegraphics[width=8cm]{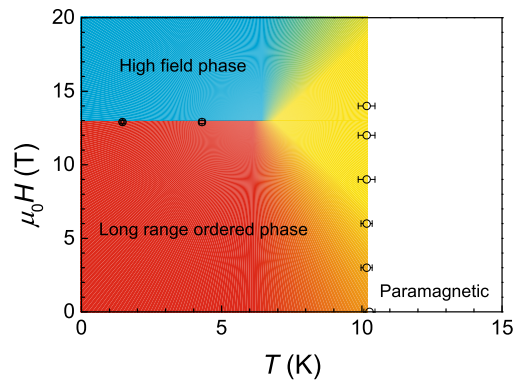}
  \caption{(Color online) The temperature-field phase diagram of NiTa$_2$O$_6$, open triangles are taken from pulse field magnetization measurements and the open squares are from specific heat measurements. The (white) area is the paramagnetic regime, the (red) area is the low field long-range order phase and the (blue) area is the high field phase whilst the yellow area is the cross-over from the low field to high field phases.}\label{phasediagram}
\end{figure}

Combining the specific heat and the pulse field magnetization results allows us to construct a temperature-field phase diagram of NiTa$_2$O$_6$ (see Figure \ref{phasediagram}). In Figure \ref{phasediagram} the white area denotes the highly correlated short-range ordered phase, the red area is the long-range ordered phase wherein we know the magnetic structure from powder neutron diffraction (see above) and the blue area is the newly discovered high field phase where the magnetic structure is not yet known. The yellow area demonstrates the broadening of the high field transition at increasing temperature. These results give the appearance of a quadratic phase diagram, which is different from other low-dimensional spin chains where the application of a field tends to suppress the ordering temperature.\cite{Banks2007,Rule2011}

\section{Conclusion}

In conclusion we investigated the magnetic properties of NiTa$_2$O$_6$ by DFT calculations, by specific heat, magnetic susceptibility, electron paramagnetic resonance, neutron powder diffraction, heat capacity  and magnetization measurements. We demonstrated that NiTa$_2$O$_6$ constitutes a $S$~=~1 Heisenberg AFM spin chain, with a nn SEI of 18.92(2)~K and a $g$-factor of 2.140(2).  This result does not support the scenario of a 2D Ising quantum AFM proposed by Santos \textit{et al.}. NiTa$_2$O$_6$ undergoes long-range AFM ordering at 10.3(1)~K and application of a magnetic field does not shift the long-range ordering anomaly. Additionally, we followed the magnetic structure as a function of temperature and determined a magnetic moment of 1.93(5)~$\mu_{\rm{B}}$ per Ni$^{2+}$ ion at 2.00(1)~K, with a critical exponent of $\beta$~=~0.22(1). The magnetic moment was found to saturate at magnetic fields larger than 55.5(5)~T with a saturation moment of 2.14(1) $\mu_{\rm{B}}$ per Ni$^{2+}$ ion at 4.3(1)~K. Lastly we presented a temperature-field phase diagram of NiTa$_2$O$_6$, wherein we mapped a new high field phase.

\section{acknowledgments}
Part of this work was funded by EuroMagNET under the EU Contract 228043.
The work at NCSU was supported by the HPC Center of NCSU for computing resources.
JML would like to thank Marc Uhlarz for experimental assistance. 

\bibliography{ref}
\bibliographystyle{apsrev}

\end{document}